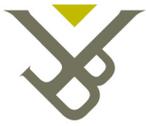

Vrije Universiteit Brussel

Universiteit Antwerpen

**Nadine Rons**[a] and **Eric Spruyt**[b]

[a] Coordinator of Research Evaluations and Policy Studies at the Research & Development Department, Vrije Universiteit Brussel, B-1050 Brussels, Belgium; e-mail: nrons@vub.ac.be

[b] Head of the Research Administration Department, Universiteit Antwerpen, B-2020 Antwerp, Belgium

**Reliability and Comparability of Peer Review Results**

*In this paper peer review reliability is investigated based on peer ratings of research teams at two Belgian universities. It is found that outcomes can be substantally influenced by the different ways in which experts attribute ratings. To increase reliability of peer ratings, procedures creating a uniform reference level should be envisaged. One should at least check for signs of low reliability, which can be obtained from an analysis of the outcomes of the peer evaluation itself.*

*The peer review results are compared to outcomes from a citation analysis of publications by the same teams, in subject fields well covered by citation indexes. It is illustrated how, besides reliability, comparability of results depends on the nature of the indicators, on the subject area and on the intrinsic characteristics of the methods. The results further confirm what is currently considered as good practice: the presentation of results for not one but for a series of indicators.*

### I. General context & research questions

While peer review is generally accepted as the principal method to evaluate research quality, its sensitivity to bias is well known. Several authors have investigated various kinds of bias in several review processes for grant and fellowship applications, manuscripts and the past performance of departments. Langfeldt (2004) e.g. distinguishes four categories of bias in peer review based on an in-depth analysis of six international evaluations of Norwegian research.

Peers may not only have different or biased opinions, they may also convert their opinions into ratings in different ways. This is important in particular when not only quality labels are given, but also relative positions or rankings are derived from the experts' ratings. The influence of an expert's bias or rating habits can be limited by including several reviews for each case being reviewed. A further possibility is asking each expert to judge more than one file, creating a uniform reference level for the ratings and discouraging some kinds of bias by making them more visible. Both mechanisms were used in the evaluations discussed in this paper. It is not always feasible however to implement procedures avoiding or correcting for biased opinions or diverging rating habits, particularly when large numbers of cases are evaluated. It is therefore important to be able to examine whether results are sufficiently reliable and take additional steps whenever necessary.

A discussion of studies on peer review reliability and on the combination of results from peer review and citation analysis is given by Moed (2005, chapter 18). At first sight more reliable results would be expected to correlate better with outcomes from other methods. Yet often relatively weak correlations are found. Observations by e.g. Aksnes & Taxt (2004) also have shown that the degree of correlation between a peer review and a citation analysis indicator depends on the scientific discipline. Explaining such discrepancies is crucial for the further acceptance of the various methodologies.

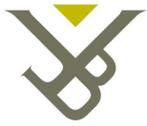

## II. Evaluation material investigated

In this paper peer ratings of research teams at two Belgian universities are investigated and compared to outcomes from a citation analysis of publications by these teams for fields well covered by citation indexes.

*Vrije Universiteit Brussel (VUB):*

A cycle of systematic peer review evaluations per discipline was introduced at the Vrije Universiteit Brussel in 1997. Two disciplines are evaluated each year. The evaluations are aimed at remediation wherever necessary, at the level of individual teams and at a higher policy level. Results for individual teams remain confidential in order to obtain clear and pertinent advice from the experts. Once completed, the peer review cycle will provide an overview of all the university's research teams and disciplines. While the focus is on the experts' comments, also quantitative scores are available. Both scores and comments are included in anonymous overviews informing the experts of their colleagues' opinions in order to prepare for the discussions during the site visit. The scores are also used to show the teams their relative position within the cluster of teams covering the scientific discipline.
As a response to the increased use of bibliometric data for the allocation of research funds in particular, a bibliometric study was conducted for the university in 2003 by the Centre for Science and Technology Studies (CWTS) (Visser et al., 2003). This very complete study included all disciplines judged to be sufficiently covered by the ISI-data bases (Thomson Scientific). In addition to this 'pure ISI' analysis, a 'target expanded' analysis was performed for two less well ISI-covered disciplines as a pilot study. The citation analysis was deliberately aimed at the same entities that are also evaluated in the systematic peer review cycle, making a comparative study possible. Apart from its utility at university policy level, the resulting debate within teams and faculties is seen as one of the study's most important effects.

*Universiteit Antwerpen (UA):*

Quality control and research assessment were also for the Universiteit Antwerpen from the early 90ties on a major point of concern and thus an important action point for the research council of the university. In close collaboration with the Centre for Science and Technology Studies (CWTS) consecutive bibliometric studies were performed in the disciplines that were considered to be covered in a significant manner (De Bruin et al., 1993, Van Leeuwen et al., 2001).
In 1998 the university decided to add peer review evaluation to the bibliometric approach. For the disciplines involved in the bibliometric studies, external peer review was sought in a purely written process based upon self assessment reports of the research teams to be evaluated, completed with some data sets on research input and output. So the same teams underwent a bibliometric study and a peer review evaluation process.
The research teams in other disciplines, generally spoken situated in the arts and humanities, felt less familiar with external evaluation procedures, so the Research Council decided to proceed in a transient phase to a process of internal peer review, complemented for the applied economics with a pilot bibliometric study (Visser et al., 2004).
All these studies will be regularly repeated following a scheme to be determined shortly.

*General & Flemish context:*

The two universities are located in the Flemish community in Belgium. The results of the evaluations are subject to a public report, as prescribed by Flemish law. Details per team are made available to the teams involved, who can comment and react to them. They are also known to the deans (VUB) or members of the research council (UA). In this way the evaluation process generates spin off at several levels. It justifies to society and authorities the major investments in research and development by Flemish authorities in the last decades; it opens within faculties the debate on the scientific subfields covered; it stimulates the research teams to modulate their research activities in a way to perform better and improve international visibility; it helps the research council and the university in the allocation of major internal research funding and in the ongoing process of the recognition and institutional support of centres of research excellence.

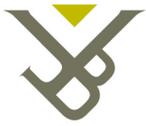

Furthermore, since the allocation amongst the universities of public research support is partly driven by research output and its quality, it is of major importance for the whole of the universities' research community to perform well on the indicators used. Since 2003 a growing part of the funding for research from the Flemish government is based on numbers of publications and citations, representing productivity and visibility. In 2005 thirty percent of the 'Special Research Fund' ('Bijzonder Onderzoeksfonds', BOF) or a total amount of 28 million euro was allocated to the Flemish universities based on publication and citation counts in the Science Citation Index Extended (SCIE). At this moment not all research areas can contribute to these publication and citation counts. In Flanders as elsewhere in the research community, an active participation is stimulated in the ongoing debate on the development of acceptable quantitative measures for the research output in the arts and humanities. Quantitative formulae for the allocation of funds at the level of universities however should not be mistaken for quality management systems. Experts warn that unwanted side effects in intra-university allocation of research funding or in publication and citation behaviours should be avoided (Debackere & Glänzel, 2004).

*Design of the peer review processes:*

The peer review evaluations examined in this paper were carefully designed to obtain pertinent and coherent results at the level of research teams. A number of principles are shared by both universities:
- Expertise-based: A review panel consists of independent and experienced experts, sufficient in number to cover the whole field (Vrije Universiteit Brussel: per discipline about as many experts as there are teams; Universiteit Antwerpen: 3 or 4 experts per discipline of 5 to 12 teams).
- International level: Experts are primarily foreign unless national experts are required by the nature of the discipline (e.g. Law). An exception was made for the less-well ISI-covered discipline evaluated at the University of Antwerp, where the panel consisted of peers chosen from within the university.
- Uniform treatment: Evaluation files and forms of uniform layout and content are presented to the experts.
- Coherence of results: All experts review all teams in a discipline, creating a uniform reference level (Vrije Universiteit Brussel: The experts evaluate all teams, or as many as they feel able to. An exception was made for one discipline where different experts were assigned to each team. Deviating individual rating behaviour is checked for. The coherence is improved by weighting each rating by the reviewer's expertise in the field, as indicated by the expert).
- Multi-criteria approach: The reviewers provide scores for different aspects related to research performance, enabling a quantitative verification of coherence and reliability. Ratings reflect the performance of teams as a whole, rather than the academic quality of individual researchers.
- Pertinent advice: Results for individual teams remain confidential, so that the teams are not refrained from presenting all relevant information and the experts feel free to formulate clear and pertinent advice.

Some additional principles are implemented by one university in particular (Vrije Universiteit Brussel):
- Sensibilization for specific research profiles: Special attention is asked for innovative and interdisciplinary research.
- Dialogue: A discussion of the experts together with the team leaders avoids misunderstandings and personal predispositions.

Table 1 gives an overview of the methods and indicators at both universities.

*Table 1: Peer review methods and indicators*

| **Vrije Universiteit Brussel (VUB)** | **Universiteit Antwerpen (UA)** | |
|---|---|---|
| - Standard procedure approved by the Research Council (Rons, 2001).<br>- The individual experts provide scores as well as comments.<br>- Weighted averages are calculated using the reviewers' expertise in each research area. | - General methodology as put forward by the VSNU (N, 1998), as well for the aspects to be judged on as for the interpretation of ratings.<br>- The individual experts provide scores.<br>- From the scores of the different experts average scores are calculated. | - Methodology specific for one discipline.<br>- The expert panel agrees upon the scores. |
| 7 disciplines,<br>5-year reference periods ('93-'97 to '98-'02) | 12 well ISI-covered disciplines,<br>7-year reference period ('92-'98) | 1 less well ISI-covered discipline,<br>8-year reference period ('92-'99) |
| *Indicators:*<br>1A: Scientific merit of the research / uniqueness of the research,<br>1B: Research approach / plan / focus / co-ordination,<br>1C: Innovation,<br>1D: Quality of the research team,<br>1E: Probability that the research objectives will be achieved,<br>1F: Research productivity,<br>2A: Potential impact on further research and on the development of applications,<br>2B: Potential for transition to or utility for the community,<br>2C: Dominant character of the research,<br>3: Reviewer's expertise in the particular research area,<br>4: Overall research evaluation. | *Indicators:*<br>A: Academic quality,<br>B: Academic productivity,<br>C: Scientific relevance,<br>D: Academic perspective. | *Indicators:*<br>a: Publications,<br>b: Projects,<br>c: Conference participations,<br>d: Other,<br>e: Globally. |
| *Scale from 1 to 10:*<br>9 to 10: High<br>7 to 8: Good<br>5 to 6: Average<br>3 to 4: Fair<br>1 to 2: Low<br>except the "Dominant character of the research": "fundamental", "applied" or "policy oriented" | *Scale from 1 to 5:*<br>5: revelatory scientific research, internationally recognised as being at the front of knowledge acquisition within the scientific discipline involved<br>4: scientific research of an international standard<br>3: scientific research resulting in a qualitatively satisfactory production with significance on a national level<br>2: scientific research on a national, non-revelatory level<br>1: insufficient level | *Scale:*<br>A: Excellent<br>B: Very good<br>C: Good<br>D: Less good |

*Bibliometric analysis:*

Two bibliometric studies (Visser et al, 2003; Van Leeuwen et al, 2001) allow to compare the peer review results to results from citation analysis in a corresponding period for the same teams in well ISI-covered disciplines. The citation analysis includes the well-established ISI-based bibliometric indicators described by Moed et al. (1995). The citation analysis indicators used in this paper include three normalized indicators, which in general would be expected to be positively correlated with peer assessments:
- CPP/JCSm: Citations per publication (CPP, excluding self citations) with respect to expectations for the journals (Mean Journal Citation Score, JCSm),
- CPP/FCSm: CPP with respect to expectations for the field (Mean Field Citation Score, FCSm),
- JCSm/FCSm: JCSm with respect to expectations for the field (FCSm).

A fourth indicator would in general be expected to be negatively correlated with peer assessments:
- PNC: Percentage of the publications that remains uncited.

**III. Reliability of results**

In this section two mechanisms influencing peer review reliability are investigated: *inter-peer agreement* and the different ways in which peers *convert opinions into ratings*. To reduce the risk of distortion of results by one specific opinion or rating habit, it is a widely accepted good practice to include 'enough' reviews for each case. Less widely practiced are procedures to obtain ratings with a more uniform reference level, such as the use of evaluation panels, with each member reviewing every file. Similar mechanisms exist in citation analysis. A particular researcher's citation behaviour can be biased by other than professional motivations (see e.g. Vinkler, 1987). The influence of such personal bias is limited as citations by many authors are included and citations to all units studied are derived from a same, large volume of citing sources.

*1. The impact of inter-peer (dis)agreement:*

Experts may have different opinions concerning a same team. When each team is rated by different experts, differences in opinion cannot be distinguished from different rating behaviours (unless also clear comments are available). When a same panel rates a series of teams however, different opinions give rise to different positions of the teams' ratings with respect to each other. The level of agreement among the experts can be measured by calculating linear correlations between their ratings for a same series of teams. This is done for the evaluations of the twelve well ISI-covered disciplines at the Universiteit Antwerpen, which are the most suitable for these calculations as every expert in the panel reviewed every team. Based on the correlation coefficients the evaluations are divided into three groups with high, intermediate and low inter-peer agreement (Table 2).

The extent to which the level of inter-peer agreement influences results can be investigated through a comparison with results from other methods. Results with better inter-peer agreement can be expected to be more reliable and therefore more likely to correspond to other assessments. Table 2 shows the linear correlations between the peer review results and the three normalized citation analysis indicators described in the previous section. Better inter-peer agreement corresponds to a higher number of significant correlations with citation analysis results. While this effect is visible at the higher aggregation level of groups of evaluations, this is not the case at the lower aggregation level of individual evaluations: results from evaluations with high inter-peer agreement may have many as well as few significant correlations with citation analysis results (Table 4). This indicates that mechanisms other than inter-peer agreement have a stronger impact on the correlation between peer review and citation analysis indicators.

Also the effect of inter-peer agreement on correlations between different peer review indicators can be expected to be limited, because a different opinion on a certain team may very well influence an expert's ratings on different aspects in a similar way. For the three groups significant correlations are found for each pair of peer review indicators, somewhat decreasing with lower inter-peer agreement. Also for the individual evaluations significant correlations are found for each pair of peer review indicators (except for two out of six pairs for one of the evaluations in the low inter-peer agreement group). Thus the correlations between different peer review indicators are found to be relatively robust for variations in inter-peer agreement.

*Table 2. Linear correlation coefficients (r) for different levels of inter-peer agreement*

| UA panel evaluations of well ISI-covered disciplines ||||||||||
|---|---|---|---|---|---|---|---|---|---|
| **a. High inter-peer agreement group** ||| **b. Intermediate inter-peer agreement group** |||| **c. Low inter-peer agreement group** |||
| 5 disciplines (Biology, Medicine 3, Physics, Chemistry, Mathematics) with: <br> - Significant inter-peer correlations at the 5% level (averages over all pairs of peers) for all indicators A to D <br> - 7 to 12 teams <br> - 3 to 4 experts per panel ||| 4 disciplines (Pharmacy, Biochemistry, Medicine 2 & 5) with: <br> - Significant inter-peer correlations at the 5% level (averages over all pairs of peers) for only some indicators <br> - 6 to 9 teams <br> - 3 to 4 experts per panel |||| 3 disciplines (Medicine 1 & 4, Informatics) with: <br> - No significant inter-peer correlation at the 5% level (averages over all pairs of peers) for any indicator <br> - 5 to 9 teams <br> - 3 experts per panel |||
| *Correlations between peer review and citation analysis indicators* ||||||||||
| - Significant correlations for all pairs of peer review and citation analysis indicators: ||| - Significant correlations for 2 out of 3 citation analysis indicators with all peer review indicators: |||| - Significant correlations for 3 out of 12 pairs of peer review and citation analysis indicators: |||
| r (N=46) | **CPP/ JCSm** | **CPP/ FCSm** | **JCSm/ FCSm** | r (N=29) | CPP/ JCSm | **CPP/ FCSm** | **JCSm/ FCSm** | r (N=18) | CPP/ JCSm | CPP/ FCSm | JCSm/ FCSm |
| A | **0,49** | **0,54** | **0,41** | A | 0,13 | **0,38** | **0,41** | A | 0,17 | **0,45** | **0,50** |
| B | **0,45** | **0,48** | **0,29** | B | 0,11 | **0,31** | **0,33** | B | 0,04 | 0,34 | **0,44** |
| C | **0,46** | **0,50** | **0,35** | C | 0,12 | **0,44** | **0,47** | C | 0,10 | 0,29 | 0,29 |
| D | **0,46** | **0,44** | **0,27** | D | -0,03 | **0,32** | **0,46** | D | 0,12 | 0,28 | 0,21 |
| *Correlations between different peer review indicators* ||||||||||
| - Significant correlations for all pairs of peer review indicators (r = **0,93** to **0,96**; N=48) ||| - Significant correlations for all pairs of peer review indicators (r = **0,89** to **0,96**; N=29) |||| - Significant correlations for all pairs of peer review indicators (r = **0,65** to **0,89**; N=21) |||
| Peer review indicators: see Table 1. <br> N = number of teams. <br> Significant correlations at the 5% level in bold. ||||||||||

*2. The impact of different rating habits:*

Experts convert their opinions into ratings in different ways, according to their own habits, corresponding to reference levels applied in other evaluations in their domain, fine-tuned in relation to scores given to other evaluated files, influenced by what the scores are believed to be used for, subject to bias, ... Often not so much the individual result, but the position with respect to other results is important. Yet when all files are reviewed independently by different experts, the experts cannot relate their ratings to those given to other files. The resulting series of scores will not necessarily be in line with the experts' opinions or comments. Having all files (or groups of files) reviewed by a same panel creates a uniform reference level and generates a series of scores that is more coherent with the experts' opinions.

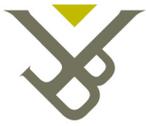

A panel-procedure was followed for all evaluations examined in this paper, except for one discipline at the Vrije Universiteit Brussel where exceptionally different experts were assigned to each team. This exception is used to illustrate how diverging rating habits influence the results and how their presence can be recognized.

For the evaluation with different experts for each team, a low correlation is observed between scores on different aspects that are expected to be correlated (Table 3, c). This particular evaluation clearly distinguishes itself hereby from the 'panel-evaluations' (Table 3, a & b), where all correlations between peer review indicators are significant. Also for the individual panel-evaluations significant correlations are found for each pair of peer review indicators (with only few exceptions). In the peer review evaluations at the Vrije Universiteit Brussel and the Universiteit Antwerpen many of the indicators can indeed be expected to be correlated to some extent. They concern global aspects of the research activities as a whole (quality, productivity, co-ordination, relevance, ...). This is not so for the evaluation of the less well ISI-covered discipline at the Universiteit Antwerpen, where the indicators mainly concern specific parts of the research activities (publications, projects, conferences, ...), which are less likely to be correlated. For the different pairs of these partial indicators no significant correlations are found (Table 3, d).

The relation between the correlation of indicators and diverging rating habits can be easily understood by examining a simplified situation. Suppose a number of fully equivalent teams are recognized by all experts as equal. The experts only convert their opinions into ratings in different ways and possibly differently for different aspects. When all teams are reviewed by a same panel, the final scores (e.g. plain averages of the individual experts' ratings) will be the same for all teams. The scores on different aspects will correlate perfectly as they are the same for each team. When on the contrary each team is reviewed by different experts, the translation of opinions into ratings will be done in a different way for each team, according to the different experts' habits. The final scores will be different for each team and the correlation between final scores on different aspects will be lower. Thus the presence of diverging rating habits has an effect on correlations between peer review indicators. A low observed correlation between aspects which are expected to be correlated (based on logic or experience) may indicate that scores are resulting from strongly diverging rating habits and need closer examination.

From the above it is clear that the impact of diverging rating habits can be limited by using a panel-procedure. A more uniform reference level may also be obtained in other ways, e.g. through a more extensive description for each indicator of the performance level linked to each score. In the absence of such measures, the presence of diverging rating habits can be revealed by correlations between different peer review indicators.

The extent to which peer review results are influenced by diverging rating habits can be investigated by comparing results from evaluations with and without a panel procedure to results from another method. In the framework of this paper only one non-panel evaluation can be compared to the panel-evaluations. Table 3 shows a comparison of the peer review results to the four citation analysis indicators described in the previous section.
For the panel-evaluations (Table 3, a & b) significant correlations are obtained for all peer review indicators with some or all of the normalized citation analysis indicators (and vice versa). Correlations with the citation analysis indicator with a negative connotation are significantly negative for (almost) all peer review indicators. There is also a striking uniformity of correlations for different peer review indicators with a same citation analysis indicator.

The correlations for the evaluation with different experts for each team (Table 3, c) do not show such uniformity and only one correlation reaches a significant level. The findings for this case indicate that diverging rating habits can influence peer review results significantly when they are not controlled in a panel-procedure. More cases will need to be examined before more general conclusions can be drawn.

*Table 3. Linear correlation coefficients (r) for the standard panel-evaluations compared to an evaluation with different experts for each team and an evaluation using 'partial' indicators*

| a. VUB panel-evaluations | | | | | c. VUB evaluation with different experts for each team | | | | |
|---|---|---|---|---|---|---|---|---|---|
| *Correlations between different peer review indicators* | | | | | | | | | |
| - 6 disciplines (Philosophy & Letters, Pharmacy, Economics, Informatics, Law, Biotechnology) <br> - Significant correlations for all pairs of peer review indicators (r = **0,67** to **0,95**; N=53) | | | | | - 1 discipline (Chemistry) <br> - Significant correlations for three pairs of peer review indicators only (r = -0,48 to **0,98**; N=5) | | | | |
| *Correlations between peer review and citation analysis indicators* | | | | | | | | | |
| - 2 disciplines (Pharmacy, Biotechnology) <br> - Significant correlations for three out of four citation analysis indicators with (almost) all peer review indicators (two positive, one negative): | | | | | - 1 discipline (Chemistry) <br> - Significant correlation for only one of the pairs of citation analysis and peer review indicators: | | | | |
| r (N=14) | CPP/ JCSm | **CPP/ FCSm** | **JCSm/ FCSm** | PNC | r (N=5) | CPP/ JCSm | CPP/ FCSm | **JCSm/ FCSm** | PNC |
| 1A | 0,39 | **0,70** | **0,64** | **-0,65** | 1A | -0,15 | -0,34 | -0,46 | 0,02 |
| 1B | 0,32 | **0,69** | **0,76** | **-0,76** | 1B | 0,25 | 0,47 | 0,56 | -0,43 |
| 1C | **0,58** | **0,85** | **0,66** | **-0,64** | 1C | -0,67 | -0,31 | 0,12 | 0,18 |
| 1D | 0,39 | **0,75** | **0,76** | **-0,60** | 1D | 0,27 | 0,67 | **0,89** | -0,60 |
| 1E | 0,31 | **0,67** | **0,74** | **-0,68** | 1E | -0,56 | -0,21 | 0,20 | 0,16 |
| 1F | 0,45 | **0,79** | **0,78** | **-0,60** | 1F | -0,06 | 0,09 | 0,15 | -0,75 |
| 2A | **0,59** | **0,80** | **0,52** | **-0,60** | 2A | -0,76 | -0,43 | -0,01 | 0,03 |
| 2B | 0,40 | **0,69** | **0,66** | -0,44 | 2B | -0,32 | 0,15 | 0,58 | -0,15 |
| 4 | 0,39 | **0,74** | **0,74** | **-0,66** | 4 | -0,42 | -0,49 | -0,50 | -0,42 |
| **b. UA panel-evaluations of well ISI-covered disciplines** | | | | | **d. UA panel-evaluation of one less well ISI-covered discipline, using 'partial' indicators** | | | | |
| - 12 disciplines (Pharmacy, Biology, Biochemistry, Medicine 1 to 5, Informatics, Physics, Chemistry, Mathematics) | | | | | - 1 discipline (Applied Economics) | | | | |
| *Correlations between different peer review indicators* | | | | | | | | | |
| - Significant correlations for all pairs of peer review indicators (r = **0,90** to **0,95**; N=98) | | | | | - No significant correlations for any of the pairs of partial peer review indicators 'publications', 'projects' and 'conference participations' (r = 0,00 to 0,53; N=10) <br> - Significant correlations for the 'global' peer evaluation with each of the partial peer review indicators (r = **0,56** to **0,63**; N=10) | | | | |
| *Correlations between peer review and citation analysis indicators* | | | | | | | | | |
| - Significant correlations for all pairs of peer review and citation analysis indicators (three positive, one negative): | | | | | | | | | |
| r (N=93) | **CPP/ JCSm** | **CPP/ FCSm** | **JCSm/ FCSm** | PNC | | | | | |
| A | **0,28** | **0,45** | **0,41** | **-0,27** | | | | | |
| B | **0,25** | **0,39** | **0,32** | **-0,22** | | | | | |
| C | **0,28** | **0,46** | **0,39** | **-0,30** | | | | | |
| D | **0,24** | **0,38** | **0,32** | **-0,24** | | Peer review indicators: see Table 1. <br> N = number of teams. <br> Significant correlations at the 5% level in bold. | | | |

**IV. Comparability with results from other methods**

One evaluation method can be used as an investigative tool for the other — as is done in this article — or as a supporting tool, monitoring tool or validation instrument. When doing so one should evidently be aware of factors potentially influencing the reliability of both types of results. Yet even when results from two methods are perfectly reliable, a comparison still is not straightforward. Two types of factors influencing comparability are demonstrated in this section: *relatedness of the indicators* and *intrinsic characteristics of the methods*.

*1. Relatedness of the indicators:*

Logically indicators describing more strongly related aspects can be expected to give rise to more similar results. This is true for comparisons within as well as between methods. The extent to which different indicators can be related is found to depend on their *global or partial nature* and on their *relevance in a particular subject area*:

- Global or partial nature of the indicator:

Correlations between peer review scores on different partial research activities (publications, projects, conferences, ...) were found to be not significant (Table 3, d), in contrast to correlations between scores on different aspects of the research as a whole (Table 3, a & b). This illustrates that performances in different kinds of research activities within a scientific discipline are not necessarily correlated. Several 'partial' peer review indicators are needed to get a complete picture. In the same way other indicators will be needed next to citation analysis indicators — related to the 'publications'-part of research activities — in order to get a complete picture of research performance.

- Relevance of the indicator for an evaluation in a particular subject area:

Correlations between a same pair of peer review and citation analysis indicators can be quite different for different disciplines. A possible explanation is that the relevance of a particular citation analysis indicator in relation to quality as evaluated by experts depends on the subject area. In this paper a high uniformity was found in the correlation coefficients for all pairs of a same citation analysis indicator and each of the peer review scores. However, for each citation analysis indicator the level of correlation with the peer review results strongly varies with the discipline (Table 4). For instance, the correlations between the peer review indicators and the citation analysis indicator reflecting journal choice (JCSm/FCSm) may be significant for one discipline and even negative for another. A possible explanation is that the experts' judgment on journal choice depends on a 'reference journal package' they have in their minds and which may be different from the reference package in the bibliometric field. This finding illustrates that one single citation analysis indicator or one constructed composite indicator may not represent a same concept for all subject areas, even among the well ISI-covered disciplines. This confirms warnings from experts in citation analysis: always more than one indicator should be presented (see e.g. Van Leeuwen et al., 2003) as composite indicators may be easily manipulated to favour part of a sample evaluated (see e.g. Grupp & Mogee, 2004).

In general and in support of earlier claims by e.g. Tijssen (2003), the findings show that any type of quantitative evaluation should be based on a series of indicators and never on one single indicator, while judgments by peers remain important for the interpretation of such quantitative results.

*Table 4. Linear correlation coefficients (r) between peer review and citation analysis indicators for the teams within a scientific discipline: variation over disciplines*

| a. VUB panel-evaluations | | | | | | | |
|---|---|---|---|---|---|---|---|
| - Pharmacy: Significant correlations for the citation analysis indicators **CPP/JCSm** and **CPP/FCSm** with six out of nine peer review indicators: | | | | - Biotechnology: Significant correlations for the citation analysis indicators **CPP/FCSm** and **JCSm/FCSm** with all peer review indicators: | | | |
| r (N=4) | **CPP/ JCSm** | **CPP/ FCSm** | JCSm/ FCSm | r (N=10) | CPP/ JCSm | **CPP/ FCSm** | **JCSm/ FCSm** |
| 1A | **0,91** | 0,86 | -0,31 | 1A | 0,08 | **0,66** | **0,88** |
| 1B | 0,88 | **0,93** | 0,41 | 1B | 0,11 | **0,68** | **0,88** |
| 1C | **0,93** | **0,96** | 0,18 | 1C | 0,40 | **0,86** | **0,85** |
| 1D | 0,88 | 0,86 | -0,19 | 1D | 0,22 | **0,75** | **0,89** |
| 1E | **0,93** | **0,92** | -0,05 | 1E | 0,07 | **0,68** | **0,94** |
| 1F | **0,94** | **0,92** | -0,16 | 1F | 0,31 | **0,81** | **0,89** |
| 2A | **1,00** | **0,99** | -0,08 | 2A | 0,31 | **0,76** | **0,79** |
| 2B | 0,88 | 0,81 | -0,50 | 2B | 0,45 | **0,81** | **0,72** |
| 4 | **0,99** | **0,97** | -0,12 | 4 | 0,17 | **0,72** | **0,88** |
| b. UA panel-evaluations, group with good inter-peer agreement | | | | | | | |
| - Biology: Significant correlations for the citation analysis indicators **CPP/JCSm**, **CPP/FCSm** and **JCSm/FCSm** with (almost) all peer review indicators: | | | | - Physics: Significant correlations for the citation analysis indicators **CPP/JCSm** and **CPP/FCSm** with all peer review indicators: | | | |
| r (N=11) | **CPP/ JCSm** | **CPP/ FCSm** | **JCSm/ FCSm** | r (N=11) | **CPP/ JCSm** | **CPP/ FCSm** | JCSm/ FCSm |
| A | **0,70** | **0,79** | **0,75** | A | **0,55** | **0,60** | 0,39 |
| B | **0,58** | **0,70** | **0,80** | B | **0,64** | **0,65** | 0,29 |
| C | **0,57** | **0,70** | **0,79** | C | **0,69** | **0,71** | 0,34 |
| D | 0,50 | **0,64** | **0,81** | D | **0,70** | **0,64** | 0,21 |
| - Medicine 3: Significant correlations for the citation analysis indicator **JCSm/FCSm** with two out of four peer review indicators: | | | | - Chemistry: Significant correlations for the citation analysis indicator **JCSm/FCSm** with two out of four peer review indicators: | | | |
| r (N=9) | CPP/ JCSm | CPP/ FCSm | **JCSm/ FCSm** | r (N=9) | CPP/ JCSm | CPP/ FCSm | **JCSm/ FCSm** |
| A | 0,39 | 0,55 | **0,68** | A | 0,05 | 0,49 | **0,67** |
| B | 0,32 | 0,51 | **0,65** | B | 0,10 | 0,34 | 0,41 |
| C | 0,18 | 0,34 | 0,54 | C | -0,03 | 0,38 | **0,58** |
| D | 0,22 | 0,36 | 0,52 | D | 0,07 | 0,32 | 0,42 |
| - Mathematics: Significant correlations for the citation analysis indicator **CPP/JCSm** with one out of four peer review indicators: | | | | | | | |
| r (N=6) | **CPP/ JCSm** | CPP/ FCSm | JCSm/ FCSm | | | | |
| A | 0,63 | 0,45 | -0,50 | | | | |
| B | 0,63 | 0,49 | -0,41 | | | | |
| C | **0,79** | 0,64 | -0,27 | | | | |
| D | 0,67 | 0,44 | -0,60 | | | | |

Peer review indicators: see Table 1.
N = number of teams.
Significant correlations at the 5% level in bold.

*2. Intrinsic characteristics of the methods:*

Even when two indicators from different evaluation methods are meant to describe a same concept, a good correlation between both is not always straightforward. Contributions to the scores may be taken into account in different ways by each method:

- Aspects intrinsic to peer review:

In the frequency distributions for several peer review indicators, a broader extension is observed towards the low scores end (e.g. for 'Research productivity', Figure 1). An extension in this direction is not observed for citation analysis results. A possible explanation is that experts may find negative elements in an evaluation file and adapt their scores accordingly (e.g. "publications of good quality but fewer than expected from the number of staff"). Standard citation analysis indicators on the other hand count all publications and citations as positive elements and give an appreciation of the publications in the database used, often regardless of the number of staff that produced them.

- Aspects intrinsic to citation analysis:

In the frequency distributions for several citation analysis indicators, a broader extension is observed towards the high scores end (Figure 2). An extension in this direction is not observed for peer review results. A possible explanation is that one or a few highly cited publications may very strongly determine a team's citation impact, more than it would influence its ratings by experts.

The available scale clearly plays a role in both cases:
- Limited to a maximum for peer review, but no maximum for the three normalized citation analysis indicators,
- In both cases limited to a minimum, but to one lying further from the mean value for the peer review indicators.

Yet scale characteristics alone cannot explain the observations, as clearly not all higher/lower end citation analysis scores correspond to higher/lower end peer review scores and vice versa (Figure 3).

In Figure 3 the teams situated in both extensions are separated from the main group of teams by a line at one standard deviation from the mean values. Most but not all of the teams in the 'high field impact'-extension (top half of the figure) also receive high peer ratings. Most but not all of the teams in the 'low peer ratings'-extension (left half of the figure) also have low field impacts.

As shown by these two examples of intrinsic characteristics of methods, the group within a studied population that is affected may be different for each method and the effect on scores may be either positive or negative. Good correlations will only be obtained when such effects can be filtered out or when they are not strongly present.

*Figure 1: Relative frequency distribution of peer results*
(VUB results, all teams in the peer review evaluations, 7 disciplines)

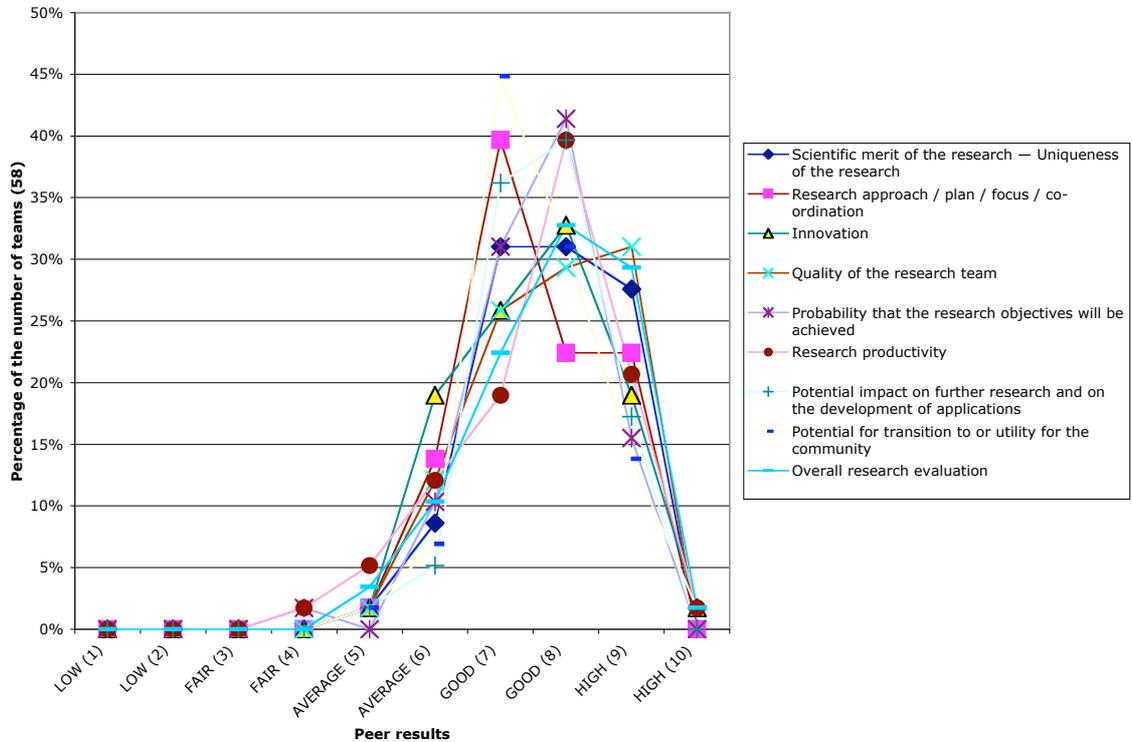

*Figure 2: Relative frequency distribution of citation analysis results*
(VUB results, all teams in the 'pure ISI' bibliometric analysis, 7 disciplines)

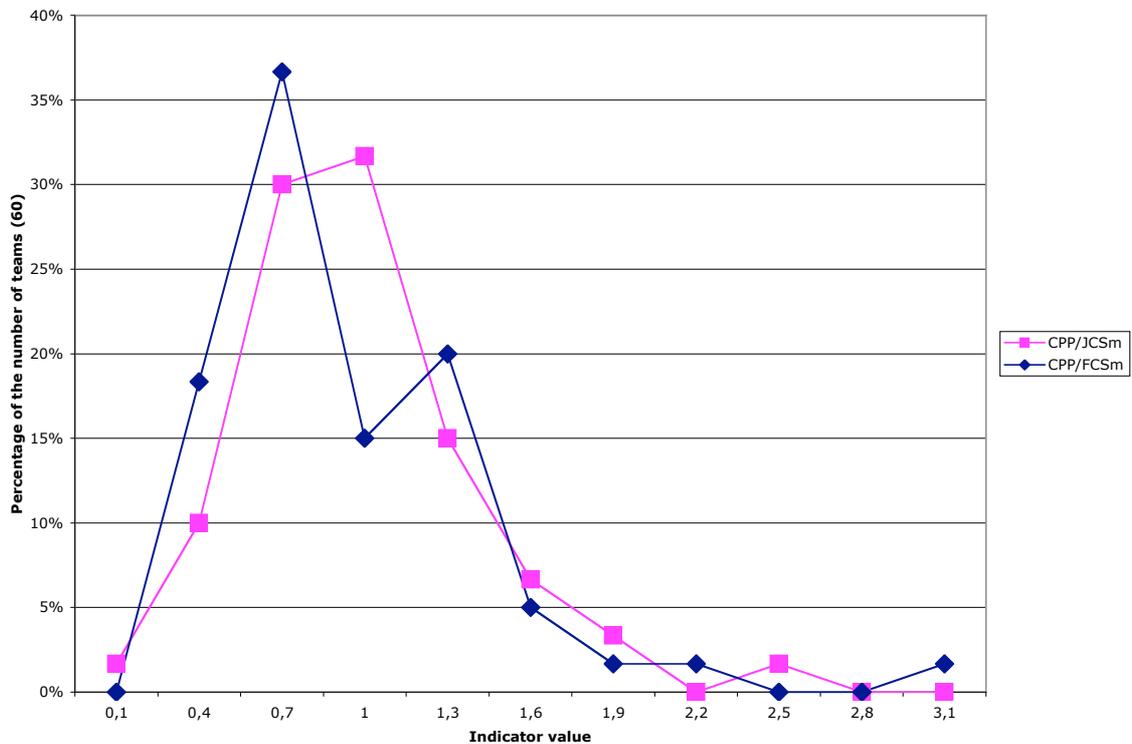

*Figure 3: Comparison of peer review and citation analysis results*
(UA results, high & intermediate inter-peer agreement groups)
Separation of the extensions towards the high scores end for CPP/FCSm and towards the low scores end for "Scientific relevance" by the horizontal & vertical lines at one standard deviation from the mean value.

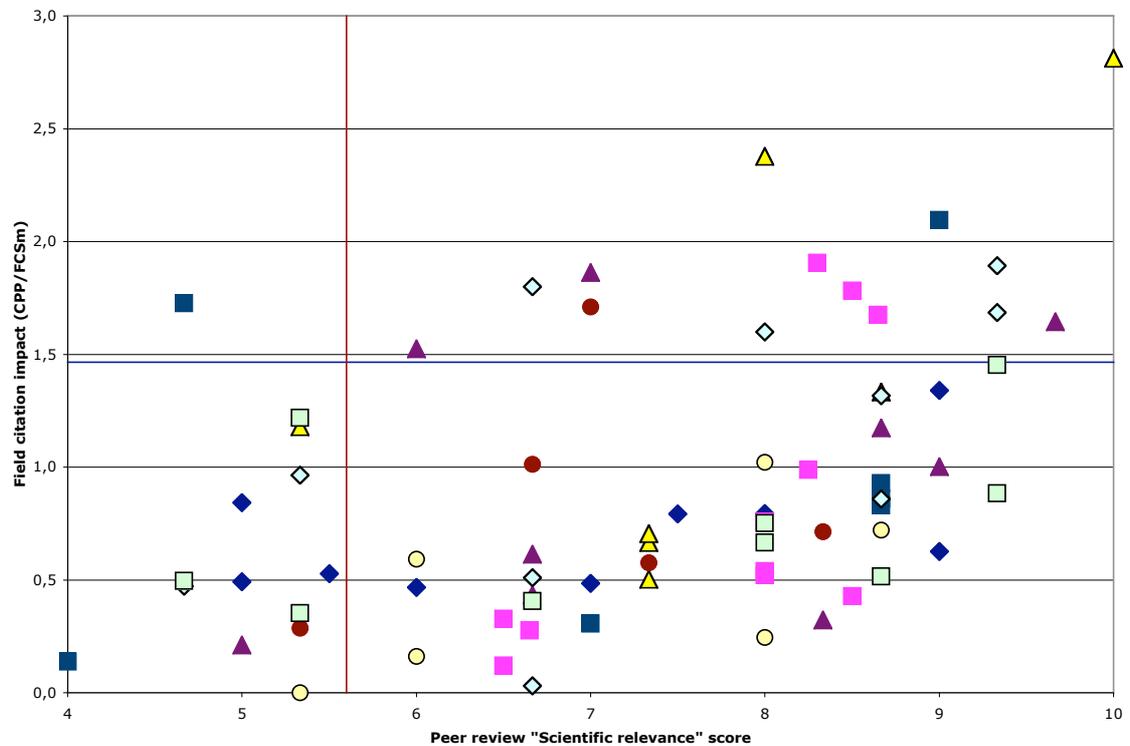

## V. Conclusions

Peer review results can be influenced considerably by the different ways in which experts convert their opinions into ratings when there is no mechanism creating a uniform reference level. For the evaluations investigated in this paper diverging rating habits appeared to have a bigger influence on results than the level of inter-peer agreement. To increase reliability of peer ratings it can be promoted as good practice to adopt procedures creating a uniform reference level. One should at least check for signs of low reliability, which can be found in correlations between different indicators of the peer evaluation itself.

Besides reliability, the comparability of results from different methods also depends on the nature of the indicators, on the subject area and on intrinsic characteristics of the methods. A weak correlation between outcomes from different methods does not imply that one method's results are 'wrong'. The methods may simply, by design, describe different aspects or emphasize different elements. In each assessment, the method or combination of methods most suitable for its purpose should be carefully chosen or developed. As a prerequisite, objectives of the review process should be clearly stated in advance.

Finally it is clear that quantitative evaluations should always be based on a series of indicators, never on one single indicator.

## Acknowledgements

The authors thank Professor Dr. Henk F. Moed (CWTS) for his valuable comments on earlier drafts leading to this paper, in particular concerning the comparisons that were made to bibliometric indicators.